\documentclass[10pt]{article}

\usepackage[T1]{fontenc}
\usepackage[utf8]{inputenc}
\usepackage[english]{babel}
\usepackage{amsmath,latexsym,amsfonts,amsthm,bm,mathtools}
\usepackage{amssymb}
\usepackage{graphicx}
\usepackage[top=1.5in, bottom=1.5in, left=1.25in, right=1.25in]{geometry}
\usepackage{booktabs,url,tikz}
\usepackage[makeroom]{cancel}
\usepackage{dsfont}
\usepackage{relsize}
\usepackage{multirow}

\usepackage{algorithm}
\usepackage{algpseudocodex}
\usepackage{bbm}
\usepackage{subcaption}

\theoremstyle{definition}

\theoremstyle{plain}

\theoremstyle{remark}

\title{Towards a Geometric Characterization of Multiverse Analysis}
\author{Giovanni Saraceno$^{\ast}$ and Antonio Calcagn\`{i} \\\\
		\footnotesize{\sl University of Padova} \\
		\footnotesize{$\ast$ E-mail: giovanni.saraceno@unipd.it}
	}

\date{}

\begin{document}

\maketitle

\begin{abstract}
Multiverse analysis makes explicit how empirical conclusions depend on alternative, defensible analytical specifications. Standard approaches usually generate the multiverse first and then summarize it through decision tables, specification curves, model weights, or scalar outputs such as estimates and \textit{p}-values. This stagewise view is useful, but it can hide how inferential uncertainty is arranged across specifications. We propose a distributional-geometric framework in which each admissible specification is represented by a probability distribution on a common target-output space. After defining a suitable distance between these distributions, the induced geometry allows the multiverse of analyses to be studied through local neighbourhoods, diameters, Fr\'echet barycentres, and dispersion measures. Numerical examples alongside a real case study illustrate how the approach complements existing multiverse summaries by retaining both effect variation and uncertainty variation.\\

\end{abstract}

\vspace{2cm}

\section{Introduction}\label{sec1}

Working with real datasets rarely requires a simple and straight path. Even when the scientific question is fixed, translating it into a statistical analysis requires a sequence of choices concerning data preprocessing, variable construction, exclusion rules, missing-data handling, covariate adjustment, model specification, and inferential summaries. In many applications, more than one option is defensible at each step, so that a single reported analysis represents only one path within a broader set of admissible analytical pipelines. The uncertainty attached to the result is therefore not exhausted by sampling variability conditional on the selected specification. Rather it also includes the variability induced by the alternative specifications of the analysis. Multiverse analysis makes this latter source of uncertainty explicit by evaluating the same research question across a structured set of analytical paths and by examining the resulting collection of statistical outputs \cite{steegen2016,simonsohn2020,girardi2024,short2026}.

The existing literature may be read as developing along two complementary directions. A first line of work is mainly descriptive and representational. Starting from the original formulation of multiverse analysis as a way to increase transparency about researcher degrees of freedom \cite{steegen2016}, this line includes specification curve analysis, where the set of reasonable specifications is defined explicitly, the corresponding estimates are displayed in an ordered curve, and the analytical decisions associated with each result are shown jointly \cite{simonsohn2020}. Related contributions have further emphasized the interpretation of multiverse analyses as collections of analytical pipelines, as well as the need for visual tools able to connect results with the decisions that generated them \cite{short2026,sarma2024}. This descriptive use of multiverse analysis has been adopted in several applied domains, including the assessment of robustness in mediation analysis \cite{rijnhart2022}, the estimation of excess mortality under alternative baseline and projection windows \cite{levitt2023}, and the reanalysis of empirical claims such as the hurricane-name example discussed in specification curve analysis \cite{simonsohn2020}. A second line of work is more explicitly inferential or evaluative. In this perspective, \cite{cantone2024} study the characterisation and calibration of multiversal methods through weighting and model-averaging schemes; \cite{girardi2024} propose post-selection inference for multiverse analysis, with global and specification-level inferential statements under multiplicity control; \cite{riha2024} integrate multiverse analysis into Bayesian modelling workflows by iteratively filtering candidate models through predictive assessment, posterior predictive checks, and computational diagnostics.

Against this background, a complementary issue concerns the statistical representation of the multiverse itself. Indeed, most existing multiverse analyses are conducted using a stagewise approach, where the multiverse is first generated and subsequently summarized in terms of specifications (e.g., decision tables, specification curves, model weights) or scalar outputs (e.g., estimates, p-values, intervals). While effective, this strategy separates the representation step (in which the multiverse is mathematically defined) from the summary step (in which it is aggregated), which inevitably discards the rich dependence structure shared across the different specifications. Motivated by the search for an end-to-end unified modeling approach, we elaborate a distributional geometric framework in which the admissible specifications are studied in terms of probability distributions and geometrically meaningful measures, such as distances, neighbourhoods, dispersion, and barycenters. Such a characterisation makes visible not only the range of possible conclusions, but also their configuration, for instance by assessing whether the multiverse is homogeneous or fragmented, whether some specifications form local clusters, and whether a proposed summary is central or peripheral with respect to the overall set of outputs.

To this end, the methodological advancement of this work unfolds along two interconnected dimensions. First, it formalizes a multiverse as a finite collection of specification-indexed probability distributions defined on a common target-output space, thereby extending the usual point-valued representation of multiverse results. Second, it introduces a geometry on the admissible specification set via distances between these inferential distributions, which in turn allows for the use of easy-to-interpret geometric measures to evaluate the multiverse of analyses. To the best of our knowledge, this is the first time multiverse analysis is characterized through a geometric approach. Consequently, the present work deliberately operates within a descriptive-statistical context, leaving asymptotic and inferential developments \cite{calin2014} to future research. 

The remainder of the paper is organised as follows. Section~\ref{sec2} introduces the proposed distributional-geometric representation of a multiverse and defines the main geometric summaries used throughout the paper. Section~\ref{sec3} discusses the operational construction of the specification-wise distributions by means of two controlled numerical examples illustrating the proposed approach. Section~\ref{sec4} applies the approach to a real multiverse analysis based on the hurricane-name data. Section~\ref{sec5} concludes with a discussion of the scope, limitations, and possible extensions of the proposed framework. The data and algorithms supporting this research study are openly available at \texttt{https://github.com/giovsaraceno/geometric-multiverse}.

\section{A distributional-geometric multiverse analysis}\label{sec2}

We now introduce the formal setting used in the rest of the paper. As in other formal treatments of multiverse analysis \cite{girardi2024}, we start from a finite set of specifications, usually elicited by the researcher before the analysis is carried out. For the sake of simplicity, we treat this set as fixed rather than stochastic.\footnote{We do not assume that the specifications are randomly sampled from a larger population of possible analyses, nor that their empirical frequencies have a direct probabilistic interpretation. Although weighted or probabilistic extensions are possible, they are not needed for the present construction.} For the reader's convenience, all notations and symbols used in this section are collected in Table~\ref{tab1}.

\subsection{Specifications, targets, and multiverse outputs}\label{sec2.1}

Let $\mathbf y \in \mathcal Y^n$ denote the observed data and let $\Sigma$ be a space of possible analytical specifications. An element $\sigma\in\Sigma$ represents a complete analytical path, including data-processing rules, variable definitions, exclusion criteria, model choices, and estimation procedures needed to address the scientific question.\footnote{In concrete applications, the relevant portion of $\Sigma$ is typically elicited from substantive and statistical expertise. This elicitation may be represented by a DAG, or by an equivalent decision structure, whose nodes encode analytical decisions and whose admissible paths define candidate specifications.} In practice, the analyst does not work with the whole space $\Sigma$, but with a finite multiverse
$$
\Sigma^\ast=\{\sigma_1,\ldots,\sigma_K\}\subseteq\Sigma,
$$
whose elements are the specifications considered admissible for the analysis. Note that $\Sigma^\ast$ need not coincide with a full Cartesian product of all analytical decisions, since some combinations may be statistically invalid, substantively unjustified, redundant, or otherwise irrelevant for the target question.

A basic requirement for the proposed construction is that all specifications in $\Sigma^\ast$ refer to a common target quantity $\boldsymbol\theta_T \in \Theta_T\subseteq\mathbb R^q $, or at least to quantities that are comparable, and can be represented on a common target-output space. This condition is essential since otherwise distances among the resulting distributions may reflect changes in the estimand itself rather than uncertainty induced by alternative analytical choices.
We denote the corresponding point-valued output by
$$
T_\sigma(\mathbf y)\in\Theta_T,
\qquad \sigma\in\Sigma^\ast.
$$
The usual point-valued multiverse can therefore be represented as the finite collection
$$
\mathcal T(\mathbf y)
=
\{T_\sigma(\mathbf y):\sigma\in\Sigma^\ast\}.
$$
In the simplest scalar case, a point-valued multiverse can be summarised through its range,
$$
R_T(\mathbf y)
=
\max_{\sigma\in\Sigma^\ast} T_\sigma(\mathbf y)
-
\min_{\sigma\in\Sigma^\ast} T_\sigma(\mathbf y).
$$
Although this quantity provides a first description of specification-induced variability, it does not distinguish, for instance, whether two similar point outputs are associated with different levels of uncertainty. This motivates the distributional representation introduced in the next subsection.

\subsection{From point-valued to distributional multiverses}\label{sec2.2}

The point-valued multiverse $\mathcal T(\mathbf y)$ describes how the selected output changes across specifications. For the present purpose, however, we associate each specification with a richer object. Let $\mathcal P(\Theta_T)$ denote a class of probability measures on the target-output space $\Theta_T$. For each $\sigma\in\Sigma^\ast$, we define
$$
Q_{\sigma,T} \in \mathcal P(\Theta_T)
$$
as a distributional representation of uncertainty about the target quantity $\boldsymbol\theta_T$ under specification $\sigma$. Note that the distribution $Q_{\sigma,T}$ is not a probability distribution over specifications. Thus, the variability across $\{Q_{\sigma,T}:\sigma\in\Sigma^\ast\}$ represents \textit{specification-induced variability}, whereas each $Q_{\sigma,T}$ represents the \textit{target-induced variability}. 

Finally, under our construction, the distributional multiverse is the finite collection
$$
\mathcal Q_T(\mathbf y)
=
\{Q_{\sigma,T}:\sigma\in\Sigma^\ast\}
\subset
\mathcal P(\Theta_T).
$$
Note that the point-valued multiverse can still be obtained from $\mathcal{Q}_T(y)$ by applying a functional $F$ to each specification-wise distribution, namely $\{F(Q_{\sigma,T}): \sigma \in \Sigma^\ast\}$. The functional $F$ may return, for example, a mean, median, mode, quantile, or interval endpoint. Although this reduction might be useful in some circumstances, it does not account for the information provided by each inferential distribution, such as dispersion, asymmetry and tail behaviour. By contrast, the full distributional multiverse representation retains this information and also allows one to study how the specification-wise distributions overlap, separate, or cluster. This collection of distributions is the object on which the geometric construction is based.

\subsection{Geometry of the distributional multiverse}\label{sec2.3}

Once each admissible specification $\sigma$ has been associated with an inferential distribution $Q_{\sigma,T}$ on the common target space $\boldsymbol\Theta_T$, the multiverse $\mathcal{Q}_T(\mathbf{y})$ can be studied as a finite configuration in a space of probability measures. This viewpoint is related to the general information-geometric idea that statistical models and families of probability distributions may be studied through geometric structures such as distances, metrics, and geodesics \cite{amari2000,sharp2022,quinn2023}. More formally, let
$$
d:\mathcal P(\Theta_T)\times\mathcal P(\Theta_T)\to[0,\infty)
$$
be a distance between probability measures on $\Theta_T$. The distance between two specifications is then induced by the distance between their corresponding inferential distributions:
$$
d_T(\sigma,\sigma')
=
d(Q_{\sigma,T},Q_{\sigma',T}),
\qquad \sigma,\sigma'\in\Sigma^\ast.
$$
According to the $d_T$ metric, two specifications are close when they lead to similar inferential distributions for the target quantity, not merely when their point estimates are close. This distinction is important because two specifications may have similar point-valued outputs but different uncertainty, or different point-valued outputs but strongly overlapping inferential distributions. It is also different from a distance on the specification space $\Sigma$. Indeed, two specifications may be close in terms of the analytical decisions that define them and yet induce substantially different distributions, while specifications that are distant as analytical paths may induce very similar distributions for the target.

\begin{table}[!h]
\caption{Main objects and summary measures used in the distributional-geometric representation of a multiverse.}
\label{tab1}
\begin{tabular}{llp{6cm}}
\hline
Object & Symbol & Meaning \\
\hline
Observed data & $\mathbf y$ & Array of observed data. \\
Specification space & $\Sigma$ & Space of possible analytical specifications. \\
Admissible multiverse & $\Sigma^\ast=\{\sigma_1,\ldots,\sigma_K\}$ & Finite set of specifications retained for the multiverse analysis. \\
Target quantity & $\boldsymbol\theta_T$ & Scientific target of interest, represented on the common target-output space. \\
Target-output space & $\Theta_T\subseteq\mathbb R^q$ & Common space in which all target-related outputs are represented. \\
Point-valued output & $T_\sigma(\mathbf y)$ & Output produced by specification $\sigma$ for the target quantity. \\
Point-valued multiverse & $\mathcal T(\mathbf y)$ & Finite collection $\{T_\sigma(\mathbf y):\sigma\in\Sigma^\ast\}$. \\
Inferential distribution & $Q_{\sigma,T}$ & Specification-wise distributional representation of uncertainty about $\boldsymbol\theta_T$. \\
Distributional multiverse & $\mathcal Q_T(\mathbf y)$ & Finite collection $\{Q_{\sigma,T}:\sigma\in\Sigma^\ast\}\subset\mathcal P(\Theta_T)$. \\
Distance between distributions & $d$ & Chosen distance on $\mathcal P(\Theta_T)$. \\
Output-induced dissimilarity & $d_T(\sigma,\sigma')$ & Dissimilarity between specifications induced by $d(Q_{\sigma,T},Q_{\sigma',T})$. \\
\hline
Point range & $R_T(\mathbf y)$ & Range of scalar point-valued outputs across specifications. \\
Distributional diameter & $\Delta_{T,d}(\mathbf y)$ & Largest pairwise distance between specification-wise inferential distributions. \\
Local point-output radius & $R_{\varepsilon,T}(\sigma_0;\mathbf y)$ & Largest point-output variation within an $\varepsilon$-ball around the reference inferential distribution. \\
Barycenter & $\overline Q_{T,d}$ & Weighted Fréchet barycenter of the specification-wise inferential distributions. \\
Barycentric dispersion & $V_{T,d}$ & Average squared distance from the barycenter. \\
\hline
\end{tabular}
\end{table}

\subsection{Summary measures}\label{sec2.4}

The metric representation above makes it possible to define summary measures for the distributional multiverse. These summaries are descriptive geometric quantities: they quantify how the specification-wise inferential distributions are arranged in $\mathcal P(\Theta_T)$ under the chosen distance $d$ and, consequently, their interpretation is conditional on the geometry induced by $d$.\\

\noindent\textbf{Local measures}. To locally summarize the distributional multiverse, we need to fix a reference specification $\sigma_0\in\Sigma^\ast$ (i.e., a baseline or reference analysis). Rather than defining a neighbourhood directly in the space of specifications, we define it in the space of probability measures. For $\varepsilon>0$, let
$$
B_d(Q_{\sigma_0,T},\varepsilon)
=
\{Q\in\mathcal P(\Theta_T):
d(Q,Q_{\sigma_0,T})\leq \varepsilon\}.
$$
Then, the corresponding local distributional multiverse around $Q_{\sigma_0,T}$ is
$$
\mathcal Q_{T,\varepsilon}(Q_{\sigma_0,T};\mathbf y)
=
\mathcal Q_T(\mathbf y)
\cap
B_d(Q_{\sigma_0,T},\varepsilon),
$$
which contains the inferential distributions associated with the specifications whose distributional outputs are within distance $\varepsilon$ from the reference distribution $Q_{\sigma_0,T}$.

To quantify the behaviour of the point-valued outputs in this local distributional region, define the $B_d(Q_{\sigma_0,T},\varepsilon)$-conditioned \textit{local radius} as follows:
\begin{equation}\label{sm_1}
R_{\varepsilon,T}(\sigma_0;\mathbf y)
=
\sup_{\sigma\in\Sigma^\ast:
Q_{\sigma,T}\in B_d(Q_{\sigma_0,T},\varepsilon)}
\left|T_\sigma(\mathbf y)-T_{\sigma_0}(\mathbf y)\right|.
\end{equation}

In the scalar case, this quantity measures the maximum variation of the reported output when one moves within an $\varepsilon$-neighbourhood of the reference inferential distribution. Large values of $R_{\varepsilon,T}(\sigma_0;\mathbf y)$ indicate local instability: specifications that are close in the distributional geometry may still produce substantially different point-valued outputs.\footnote{Here and below, \textit{large} is meant relative to the chosen geometry. Hence, its interpretation depends on the distance $d$ and the target-output scale, which in turn depends on the given application being considered.}\\

\noindent\textbf{Global measures}. A global geometric assessment of the distributional multiverse is given by its \textit{diameter},
\begin{equation}\label{sm_2}
\Delta_{T,d}(\mathbf y)
=
\sup_{\sigma,\sigma'\in\Sigma^\ast}
d(Q_{\sigma,T},Q_{\sigma',T}),
\end{equation}
which measures the largest distributional discrepancy induced by the multiverse. Large values of $\Delta_{T,d}(\mathbf y)$ indicate that the inferential distributions associated with the admissible specifications are highly heterogeneous under the chosen geometry.

The distributional multiverse can be also summarised through a distributional average of the specification-wise inferential distributions. Given non-negative weights $\{w_\sigma\}_{\sigma\in\Sigma^\ast}$ such that \\$\sum_{\sigma\in\Sigma^\ast}w_\sigma=1$, define a \textit{weighted Fr\'echet barycenter} by
\begin{equation}\label{sm_3}
\overline Q_{T,d} = \arg\min_{Q\in\mathcal P(\Theta_T)} \sum_{\sigma\in\Sigma^\ast} w_\sigma d^2(Q_{\sigma,T},Q),
\end{equation}
which is a geometric summary of the finite configuration $\mathcal Q_T(\mathbf y)$.
Notice that this does not correspond to a selected specification, and it should not be confused with a posterior distribution over specifications. If a point-valued summary is desired, it can be obtained by applying a suitable functional $F$ to the above barycenter,
\begin{equation}\label{sm_3b}
\widetilde T(\mathbf y)=F(\overline Q_{T,d}),
\end{equation}
where $F$ may be an expectation, median, quantile, interval functional, or another summary appropriate for the target quantity.

Finally, a \textit{Fr\'echet-type dispersion measure} is the average squared distance from the barycenter, namely
\begin{equation}\label{sm_4}
V_{T,d} = \sum_{\sigma\in\Sigma^\ast} w_\sigma d^2(Q_{\sigma,T},\overline Q_{T,d}),
\end{equation}
which measures the overall spread of the distributional multiverse around its barycentric summary. Unlike $\Delta_{T,d}(\mathbf y)$, the quantity $V_{T,d}$ depends on the whole configuration rather than only on the most distant pair of specifications.

\subsection{Remarks}\label{sec2.5}

\textbf{Remark 1} \textit{Discrete and continuous specification sets}. 
The construction above is stated for a finite multiverse $\Sigma^\ast$, which is the usual situation in applied multiverse analysis. In this case, $\Sigma^\ast$ can be regarded as a discrete grid, or as a finite lattice-like structure when the admissible analytical decisions are partially ordered. The induced object $\mathcal Q_T(\mathbf y)$ is therefore a finite configuration in $\mathcal P(\Theta_T)$. Note that this discreteness is not essential to our general idea. As a matter of fact, specifications may include continuous analytical choices, such as thresholds, tuning parameters, prior hyperparameters, or constraints ranging over intervals. In these settings, the specification set may be modelled as a subset of $\mathbb R^p$, or more generally as a smooth manifold. Particularly, if $\Sigma^\ast$ carries such a smooth structure, if $\mathcal P(\Theta_T)$ is restricted to a class of probability measures with a compatible differentiable structure, and if the map $\sigma \mapsto Q_{\sigma,T}$ is sufficiently regular, then the distributional multiverse may itself be regarded as an embedded manifold in $\mathcal P(\Theta_T)$. The finite cloud of specification-wise distributions is therefore replaced by a curve, surface, or higher-dimensional smooth object. As a result, the continuous case should be understood as a possible extension of the same geometric viewpoint, rather than as an assumption required by the proposed framework.\\

\noindent \textbf{Remark 2} \textit{On the choice of $d$}.
The choice of $d$ (e.g., Hellinger, Wasserstein, Fisher-Rao) is not merely technical: it induces the geometry in which the distributional multiverse $\mathcal Q(\mathbf y)$ is represented, and therefore fixes the sense in which two specification-wise inferential distributions, say $Q_{\sigma,T}$ and $Q_{\sigma',T}$, are regarded as close or distant. Depending on the metric, this distance may be interpreted, for example, as a geodesic length in a statistical or information-geometric space (Fisher-Rao), as an optimal-transport cost between probability measures (Wasserstein), or as a measure of discrepancy in mass, overlap, or tail behaviour. Different choices of $d$ therefore emphasize different aspects of the same collection of inferential distributions and may lead to different geometric readings of the multiverse. When no application-specific reason suggests a different metric, a natural default choice is the quadratic Wasserstein distance $D_2$, provided that the specification-wise distributions have finite second moments. This choice is simple but still effective because it compares distributions on the scale of the target quantity, admits an optimal-transport interpretation, and leads to a well-defined notion of barycenter in many standard settings.\\

\noindent \textbf{Remark 3} \textit{On the construction of \(Q_{\sigma,T}\).} There are several ways through which $Q_{\sigma,T}$ can be constructed. In a frequentist analysis, $Q_{\sigma,T}$ can be obtained from an asymptotic approximation to the sampling distribution of the estimator. For instance, if $T_\sigma(\mathbf y)$ is an estimator of $\boldsymbol\theta_T$ and $\widehat{\mathbf S}_\sigma(\mathbf y)$ is the estimated asymptotic covariance matrix, one can set
$$
Q_{\sigma,T}^{\mathrm{asy}}
=
\mathcal N_q\{T_\sigma(\mathbf y),\widehat{\mathbf S}_\sigma(\mathbf y)\}.
$$
Alternatively, $Q_{\sigma,T}$ may be constructed by bootstrap resampling. If $\mathbf y^{\ast 1},\ldots,\mathbf y^{\ast B}$ are bootstrap samples and $T_\sigma(\mathbf y^{\ast b})$ is the corresponding output under specification $\sigma$, then
$$
Q_{\sigma,T}^{\mathrm{boot}}
=
\frac{1}{B}\sum_{b=1}^B \delta_{T_\sigma(\mathbf y^{\ast b})},
$$
where $\delta_x$ denotes the Dirac measure at $x$. Finally, in a Bayesian analysis, $Q_{\sigma,T}$ may be taken as the posterior distribution of the target quantity under specification $\sigma$. When the target is obtained from model-specific parameters through a transformation, we assume that the posterior has already been mapped to the common target-output space $\Theta_T$. Thus, writing $\pi_\sigma(\boldsymbol\theta_T\mid \mathbf y)$ for this induced posterior distribution, we set $$ Q_{\sigma,T}^{\mathrm{Bayes}} = \pi_\sigma(\boldsymbol\theta_T\mid \mathbf y). $$ Note that in this case, the posterior is already expressed directly on the common target-output space.\\

\noindent \textbf{Remark 4} \textit{On the nature of \(\mathcal{Q}_T(\mathbf y)\).}
The construction proposed here is finite and descriptive. We do not aim to define a new topology on \(\mathcal P(\Theta_T)\), nor to establish general measure-theoretic properties of the space of probability measures. Once a distance \(d\) has been chosen, such properties belong to the existing theory of the corresponding metric geometry, for example optimal-transport geometry in the Wasserstein case \cite{villani2009,santambrogio2015}. In the present setting, the relevant object is the finite metric configuration obtained after the specification-wise distributions \(Q_{\sigma,T}\) have been constructed and a distance \(d\) has been selected. The resulting summaries are therefore conditional on the admissible specification set \(\Sigma^\ast\), on the construction of the inferential distributions \(Q_{\sigma,T}\), and on the chosen distance $d$. \\

\section{Numerical examples}\label{sec3}

In this section, we use controlled numerical examples to illustrate how the proposed distributional-geometric approach works in practice. 

\subsection{Example 1}\label{sec3.1}

This first example illustrates the basic workflow for running a multiverse analysis using the proposed geometric approach. For this purpose it is deliberately kept simple in the formulation. We also show that a point-wise multiverse, since it depends on scalar summaries, tends to collapse effect magnitude and uncertainty into a single number, making it difficult to distinguish specifications that differ in uncertainty only. By contrast, the distributional multiverse keeps these two components jointly represented but conceptually distinct, allowing for simple but still effective analyses of different specifications. \\

\noindent\textit{Data.} We simulated $n=500$ data from a heteroskedastic Normal linear model with one centered continuous covariate $x_{1}$ and a binary predictor $x_{2i}\in\{A,B\}$ with $P(x_{2i}=B)=0.25$. Let $z_i = \mathbbm{1}(x_{2i}=B)$. The data generating model is $y_i = \mu_i + \epsilon_i$, where $\mu_i = \beta_0 + x_{1i}\beta_1 + x_{1i}^2\beta_2 + z_{i}\beta_3 + x_{1i}x_{2i}\gamma$, and 
$\varepsilon_i | x_{2i} \sim \mathcal N(0, \sigma^2_{x_{2i}})$, with $\sigma^2_{x_{2i}}$ depending on the levels of the categorical predictor. 
In the data-generation schema, the true regression coefficients are set as follows $\beta_0 = 0.0$, $\beta_1 = 0.35$, $\beta_2 = 0.12$, $\beta_3 = -0.45$, and $\gamma = 0.06$. 
The true mean model therefore contains both a quadratic term in $x_1$ and a first-order interaction between $x_1$ and the group indicator.
The target quantity is the group contrast $\theta_T = \mathbb E[Y|x_2=B]-E[Y|x_2=A]$ evaluated at $x_1=0$, which corresponds to $\beta_3$ in the considered model. The simulated data include an unbalanced group structure (i.e., $z_{i} \sim \mathcal Ber(0.25)$) and heteroskedastic errors (i.e., $\sigma_i=0.40$ for the first group, and $\sigma_i=1.20$ for the second group). \\

\noindent\textit{Specification set.} The finite multiverse is generated by the cartesian product of three binary analytical choices
\begin{align*}
	\Sigma^* = \{\text{additive},\text{first-order interaction}\} &\times \{\text{homoskedastic},\text{heteroskedastic}\} \times & \\
	&\times \{\text{linear},\text{quadratic}\},
\end{align*}
leading to a total of $2^3 = 8$ possible specifications. The first choice concerns the mean structure, distinguishing an additive model from a model including the first-order interaction between $x_2$ and $x_1$. The second choice concerns the uncertainty estimator, distinguishing classical homoskedastic standard errors from heteroskedasticity-robust standard errors. The third choice concerns the adjustment for the continuous covariate, distinguishing a linear adjustment from a quadratic adjustment.\\

\noindent\textit{Point-valued multiverse $\mathcal T(\mathbf y)$.} For each specification $\sigma\in\Sigma^\ast$, the point-valued output is $T_\sigma(\mathbf y)=\widehat\theta_{T,\sigma} = \hat\beta_{3, \sigma}$, namely the estimated group contrast. The resulting point-valued multiverse is the finite collection $\mathcal T(\mathbf y) = \{\hat\beta_{3,\sigma}:\sigma\in\Sigma^\ast\}$. \\

\noindent\textit{Distributional multiverse $\mathcal Q_T(\mathbf y)$.} For each specification, we approximate the inferential distribution of the target by
$Q_{\sigma,T}=\mathcal N(\hat\beta_{3,\sigma},\hat{\mathrm{se}}_{\beta_{3,\sigma}}^2)$, yielding the finite collection
$\mathcal Q_T(\mathbf y) = \{\mathcal N(\hat\beta_{3,\sigma},\hat{\mathrm{se}}_{\beta_{3,\sigma}}^2):\sigma\in\Sigma^\ast\}$. Note that, unlike for the point-wise multiverse construction, 
the latter formulation is richer in terms of representation, as it retains the distribution of the effects directly into the analysis. \\

\noindent\textit{Wasserstein geometry.} Given the characteristics of $Q_{\sigma,T}$, the natural choice for defining a geometry on $\mathcal Q_T(\mathbf y)$ is the quadratic Wasserstein distance, which in case of scalar Gaussian distributions takes the simple form
$$ W_2^2(Q_{\sigma,T},Q_{\sigma',T}) = (m_\sigma-m_{\sigma'})^2 + (s_\sigma-s_{\sigma'})^2,$$ with
$m_\sigma=\hat\beta_{3,\sigma}$, and $s_\sigma=\hat{\mathrm{se}}_{\beta_{3,\sigma}}$. As a consequence, $\mathcal Q_T(\mathbf y)$ can be represented geometrically in the plane $(m_\sigma,s_\sigma)$. \\

\begin{figure}[!h]
	\centering
	\includegraphics[scale=0.65]{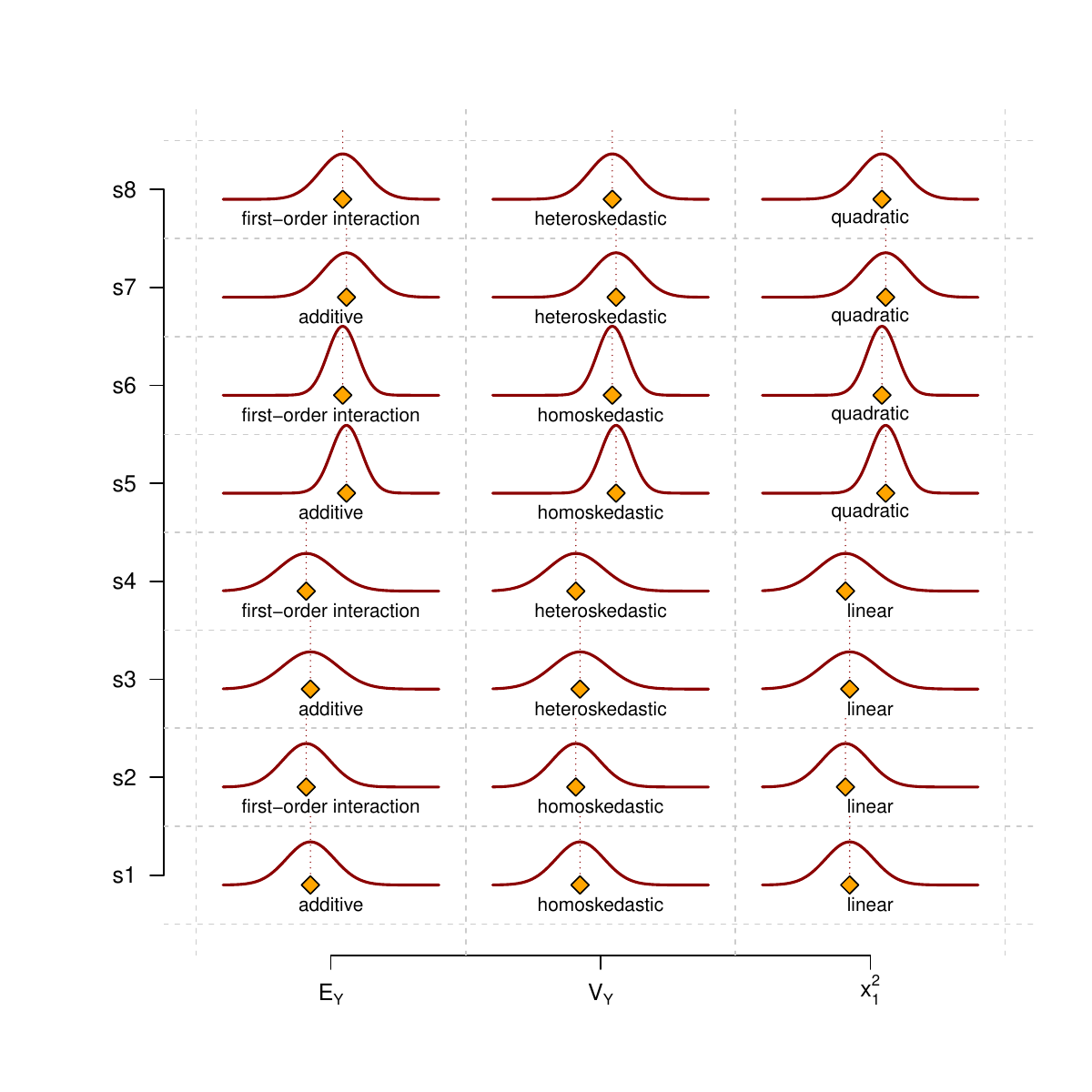}
	\caption{Numerical example 1. Specification grid, point-wise effects (coloured yellow diamonds), and distributional effects (darkred curves), for each modeling specification.}
	\label{fig1}
\end{figure}

\noindent\textit{Results.} The first controlled example illustrates how the distributional representation helps to separate specification-induced variability from target-induced variability. Figure \ref{fig1} shows the point-valued against the distributional-valued effect for each modeling specification $\sigma \in \Sigma^*$. The estimated effects mainly depend on the covariate-adjustment term. Indeed, specifications using the linear adjustment produce more negative effects, approximately between $-0.61$ and $-0.60$, whereas specifications using the quadratic adjustment produce less negative effects, approximately between $-0.45$ and $-0.43$. By contrast, the additive versus first-order interaction choice has only a minor effect on the estimated contrast within each covariate-adjustment regime, and the homoskedastic versus heteroskedastic choice does not affect the point estimate itself. Overall, the point-valued range is $R_T(\mathbf y)=0.1865$. This measure quantifies the largest absolute difference between estimated effects, but it ignores the fact that the same estimated effect may be associated with different levels of uncertainty. This is the case, for instance, of the specifications $\sigma_5$ (i.e., \{additive,homoskedastic,quadratic\}) and $\sigma_7$ (i.e., \{additive,heteroskedastic,quadratic\}), which are close in their point-valued effect but differ in the target-induced uncertainty. The distributional representation extends this comparison by replacing each point estimate with the corresponding inferential distribution $Q_{\sigma,T}$. Under the Wasserstein geometry, the distributional diameter is $\Delta_{T,W_2}(\mathbf y)=0.1950$, which is slightly larger than $R_T(\mathbf y)$. Although both quantities are absolute distances on the target scale, the former accounts for variation in both location and scale. Decomposing the squared Wasserstein distance shows that most of the discrepancy is due to the
difference in locations, while a smaller contribution is due to the difference in standard errors. Continuing with the summary measures, the unweighted Wasserstein barycenter is
$ \overline Q_{T, W_2} = \mathcal N(-0.5209,0.1041^2)$, which lies between the two covariate-adjustment regimes. Unlike the average point-valued effect, $\overline Q_{T,W_2}$ summarizes where the multiverse is geometrically centred while retaining an uncertainty component. The barycentric dispersion is
$ V_{T,W_2} = 0.0076$, which quantifies the average spread of the specification-wise inferential distributions around their Fr\'echet mean (i.e., the specification-induced variability). Taken together the Fr\'echet mean $\overline Q_{T,W_2}$ and the Fr\'echet-type variance $V_{T,W_2}$ provide a compact summary of the central tendency and heterogeneity of the distributional multiverse.\footnote{An alternative point-valued summary would be the test statistic $t_{\beta_3,\sigma}$. In the present example, the corresponding multiverse gives a different reading, because heteroskedasticity-robust specifications have less extreme test statistics than their homoskedastic counterparts even when the estimated effect is unchanged. This illustrates that point-valued multiverses depend on the scalar summary chosen: estimated effects emphasize variation in location, whereas test statistics collapse effect magnitude and uncertainty into a single number. The distributional representation avoids this collapse by retaining both components in $Q_{\sigma,T}$.}

\begin{figure}[!h]
	\centering
	\includegraphics[scale=0.5]{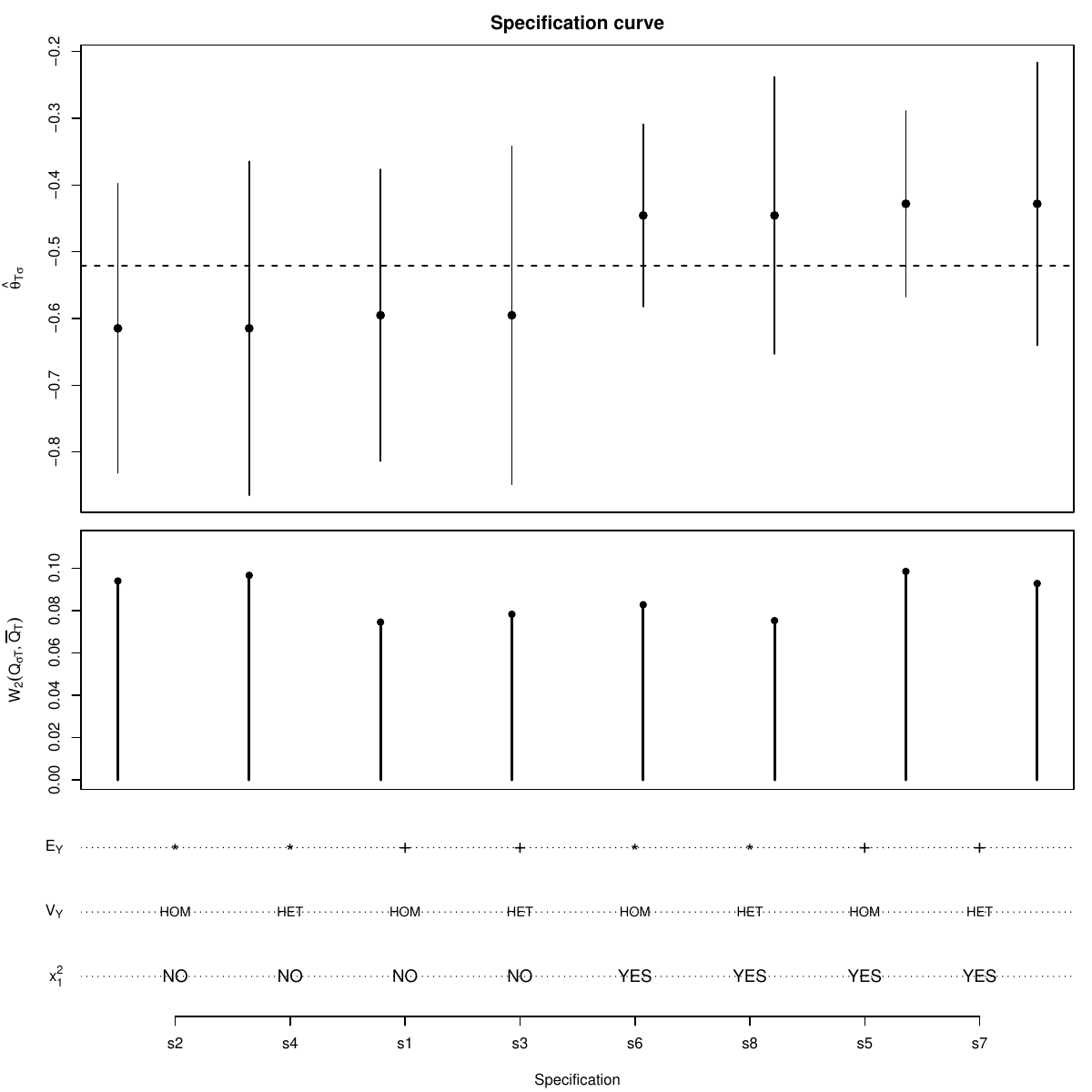}
	\caption{Numerical example 1. Specification-curve of the distributional multiverse (specifications are ordered according to the estimated effect). The upper panel reports the point-valued outputs together with the corresponding normal approximation intervals, whereas the dashed horizontal line denotes the mean of the Wasserstein barycenter. The middle panel reports the Wasserstein distance of each specification-wise inferential distribution from the barycenter. The lower panel displays the three modeling specification: mean structure (additive $+$ vs. interaction $*$), variance estimator (HETeroskedastic vs HOMoskedastic), and covariate adjustment ($x_1^2$ YES vs $x_1^2$ NO). }
	\label{fig2}
\end{figure}

Finally, Figure~\ref{fig2} provides a specification-curve-like representation of the distributional multiverse. The upper panel shows that the estimated effects are mainly separated by the covariate-adjustment choice: linear adjustment leads to more negative contrasts, whereas quadratic adjustment leads to less negative contrasts. The uncertainty intervals, however, reveal an additional source of variation: heteroskedasticity-robust specifications have larger standard errors than their homoskedastic counterparts, even when the point estimate is unchanged. The middle panel reports the Wasserstein distance of each specification-wise inferential distribution from the barycenter $\overline Q_{T,W_2}$, which summarizes the geometric position of each $Q_{\sigma,T}$ relative to the centre of the distributional multiverse (hence reflecting also differences in uncertainty). Overall, the figure shows that a standard point-valued specification curve identifies the polynomial adjustment as the main driver of effect variation, whereas the distributional specification curve also displays the contribution of the variance estimator to the geometry of the multiverse.

\subsection{Example 2}\label{sec3.2}

This example considers a setting in which the multiverse is generated by continuous tuning choices, rather than by a few categorical modeling choices. To this end, we use a simple regression-discontinuity (RD) design, where the goal is to estimate the jump in the outcome at a known cutoff of a running variable. Here, we vary two choices: the bandwidth, which determines how wide the window around the cutoff is, and the donut-hole radius, which removes observations that are extremely close to the cutoff \cite{cattaneo2020,noack2023}. Each combination of bandwidth and donut-hole radius defines one specification. Varying both quantities therefore produces a two-dimensional multiverse of RD analyses.\\

\noindent\textit{Data.} Let $x_i$ be a running variable centered at the cutoff $0$, and let $z_i=\mathbbm 1(x_i\geq 0)$ denote the treatment indicator. We simulate two outcomes with $n=200$ observations, according to $y_{ij} = \alpha_{0j}+\tau_j z_i+\beta_{j}x_i+\varepsilon_{ij}$,  $j\in\{1,2\}$. Here, $\alpha_{0j}$ is the baseline intercept for outcome $j$, $\beta_{j}$ is the slope for the running variable, and $\tau_j$ is the discontinuity at the cutoff for outcome $j$. The target is the bivariate vector of RD effects $\boldsymbol\theta_T=(\tau_1,\tau_2)^\top \in \mathbb R^2$. In the data-generation, the running variable is simulated around the cutoff, that is  $x_i\sim U(-1.5,1.5)$,  $\boldsymbol\tau=\{0.7,-0.4\}$, $\boldsymbol\alpha_0 = \{0.2,-0.1\}$, $\boldsymbol\beta=\{0.8,-0.5\}$.
The error term for the first outcome $\boldsymbol\epsilon_1$ is Gaussian with mean zero and standard deviation $\sigma_{\epsilon_1} = 0.7$. The second error term $\boldsymbol\epsilon_2$ is also Gaussian with the same marginal standard deviation but centered on $\frac{1}{2}\boldsymbol\epsilon_1$ (so that it is correlated with the first term). This produces a bivariate target. \\

\noindent\textit{Specification set.} In this example, each specification of the multiverse is generated by two continuous choices, i.e. $\sigma=(h,r)$. The first is the bandwidth $h>0$ which determines the window $|x_i| \leq h$ of observations retained for the local RD fit. The second is the donut-hole radius $r\geq0$ which excludes  observations with $|x_i|<r$ before the RD fit. Note that the bandwidth $h$ is used to perform the local-polynomial fit around the cutoff using the triangular kernel. The specification set is therefore $\Sigma=[h_{\min},h_{\max}]\times[0,r_{\max}]$, which in numerical implementation is replaced by the finite grid $\Sigma^\ast = \{(h_m,r_l):m=1,\ldots,M_h,\ l=1,\ldots,L_r\}$. In this example we consider $M_h=24$ bandwidth values and $L_r=16$ donut-hole values yielding $|\Sigma^\ast| = M_h \times L_r = 24 \times 16 = 384$ admissible specifications.\\

\noindent\textit{Point-valued multiverse $\mathcal T(\mathbf y)$.} For each specification $\sigma=(h,r)\in\Sigma^\ast$, the point-valued output is $T_\sigma(\mathbf y)=\boldsymbol\theta_{T,\sigma} = (\hat\tau_{1_{h,r}},\hat\tau_{2_{h,r}})$, namely the RD estimated effects. The point-valued multiverse is the finite collection $\mathcal T(\mathbf y) = \{(\hat\tau_{1_{h,r}},\hat\tau_{2_{h,r}}):(h,r)\in\Sigma^\ast\}$. \\

\noindent\textit{Distributional multiverse $\mathcal Q_T(\mathbf y)$.} As for the previous case, the distributional multiverse here is obtained through the Gaussian sampling distribution of the RD effect centred on $\boldsymbol{\hat\tau}_{h,r}=(\hat\tau_{1_{h,r}},\hat\tau_{2_{h,r}})$ and diagonal covariance matrix $\mathbf{\hat S}_{h,r}$ with non-zero elements equal to $(\widehat{\mathrm{se}}^2_{\tau_{1_{h,r}}},\widehat{\mathrm{se}}^2_{\tau_{2_{h,r}}})$. Hence, the distributional multiverse is the collection $\mathcal Q_T(\mathbf y) = \{\mathcal N_2(\boldsymbol{\hat\tau}_{h,r},\mathbf{\hat S}_{h,r}):(h,r)\in\Sigma^\ast\}$.\\

\noindent\textit{Hellinger geometry.} Unlike for Example 1, we used the Hellinger distance for the multiverse geometry. This type of distance measure is convenient here because it is bounded, has a closed form for Gaussian distributions, and directly measures distributional overlap, which is relevant once we deal with bivariate Gaussian distributions. In the bivariate Gaussian case, the Hellinger distance is directly computed from the Bhattacharyya coefficient (BC) as follows 
$$
H(Q_{\sigma,T},Q_{\sigma',T})
=
\left\{1-\operatorname{BC}(Q_{\sigma,T},Q_{\sigma',T})\right\}^{1/2},
$$  
where 
$$
\operatorname{BC}(Q_{\sigma,T},Q_{\sigma',T})
=
\frac{|S_{\sigma,T}|^{1/4}|S_{\sigma',T}|^{1/4}}
{|\bar S_{\sigma,\sigma',T}|^{1/2}}
\exp\left\{
-\frac{1}{8}
(\mu_{\sigma,T}-\mu_{\sigma',T})^\top
\bar S_{\sigma,\sigma',T}^{-1}
(\mu_{\sigma,T}-\mu_{\sigma',T})
\right\},
$$
and $\bar S_{\sigma,\sigma',T}=\frac{1}{2}(S_{\sigma,T}+S_{\sigma',T})$. Note that, unlike the Wasserstein distance, the Hellinger distance is bounded between $0$ and $1$ and measures lack of overlap between two bivariate distributions associated with two different specifications. Given the Hellinger-induced geometry, the distributional multiverse $\mathcal Q_T(\mathbf y)$ can be represented in terms of ellipses whereas its geometry can be conveniently represented in a lower-dimensional space via multidimensional scaling. \\

\noindent\textit{Results}. The second controlled example complements Example~1 by describing a multiverse analysis on a bivariate target, with the specification set being a dense discretization of a continuous two-dimensional domain rather than a small finite Cartesian product. Figure \ref{fig3} provides the bivariate counterpart of Figure \ref{fig1} in Example~1. Particularly, Panel A shows that the specification-wise estimates are not arranged along a single scalar direction: most specifications form a central cloud, while a smaller group moves toward lower values of $\widehat\tau_2$ and wider uncertainty ellipses. The displayed medoid ellipse is located close to the Fr\'echet barycenter whereas the two external ellipses corresponding to the Hellinger diameter represent two different ways of being distributionally far apart: one is attached to a small-bandwidth and the other to a larger-bandwidth specification with a much tighter inferential distribution. Panel B displays the same information through the geometry of the Hellinger distance matrix, here projected onto a lower-dimensional space as provided by the Multidimensional scaling. The geometry here is curved rather than flat or nearly linear, which indicates that the multiverse is better read as a local geometry of inferential distributions than as a single ordered list of point estimates. The point-valued multiverse shows that $\widehat\tau_1$ ranges from $0.2374$ to $2.1220$, whereas $\widehat\tau_2$ ranges from $-1.1502$ to $0.8389$. The Euclidean point-valued diameter in the target space, which is computed here as the maximum distance among all the possible specifications, is $R_T(\mathbf y)=2.5776$. Similarly, under the Hellinger geometry, the distributional diameter is $\Delta_{T,H}(\mathbf y)=0.9617$, attained by the specifications with $(\hat\tau_1,\hat\tau_2)_{-}=(0.37,-0.56)$ (leftmost black dot in Figure \ref{fig3}-A) and $(\hat\tau_1,\hat\tau_2)_{+}=(1.97,0.38)$ (rightmost black dot in Figure \ref{fig3}-A). These two specifications also differ sharply in uncertainty, as their standard-error vectors are $\mathrm{diag}(\mathbf{\hat S}_{-}) = (0.2470,0.2422)$ and $\mathrm{diag}(\mathbf{\hat S}_{+}) = (0.0753,0.1987)$. The unweighted Hellinger barycenter is
$$
\overline Q_{T,H}
=
\mathcal N_2
\left(
\begin{pmatrix}
1.0025\\
-0.1472
\end{pmatrix},
\begin{pmatrix}
0.1408 & 0.0277\\
0.0277 & 0.2017
\end{pmatrix}
\right),
$$
whereas the barycenter dispersion is $V_{T,H}=0.1558$. 
\begin{figure}[!h]
	\centering
	\includegraphics[scale=0.5]{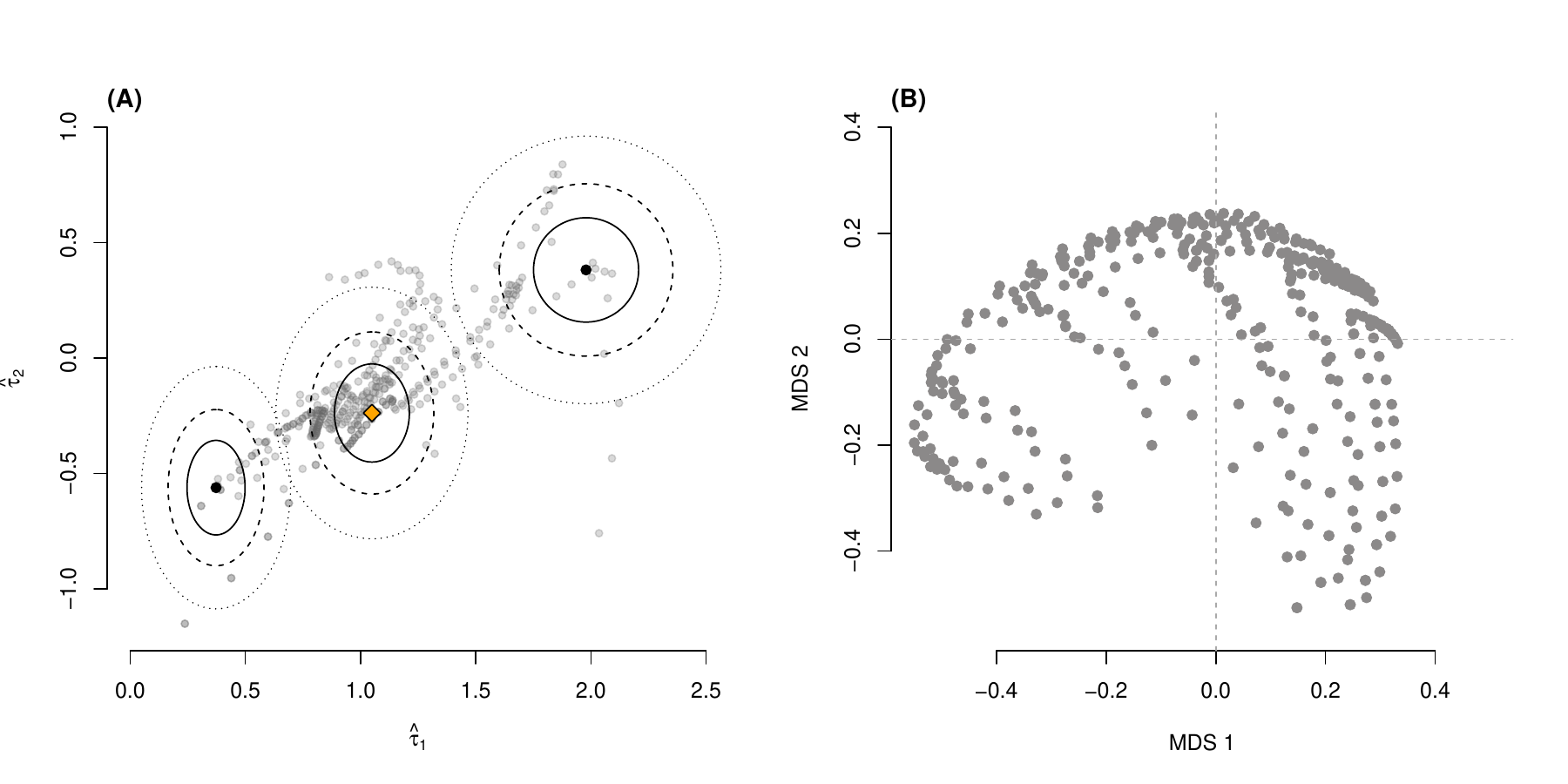}
	\caption{Numerical example 2. Panel (A): Point-wise multiverse (gray dots) and a subset of specification ellipses from the distributional multiverse (drawn at the levels 0.10, 0.25, and 0.50). Note that the yellow diamond is the medoid whereas the black dots represent the most external effects closer to the frontier of the multiverse. Panel (B): Multidimensional Scaling-based representation of the Hellinger-based geometry of $\mathcal Q_T(\mathbf y)$.}
	\label{fig3}
\end{figure}
The Hellinger distance can also be used to define local regions of the distributional multiverse through a local $\varepsilon$-neighbourhood of the barycenter $\overline{Q}_{T,H}$, namely $B_d(\overline{Q}_{T,H},\epsilon) = \left\{\sigma\in\Sigma^\ast:H(Q_{\sigma,T},\overline Q_{T,H})\leq\varepsilon\right\}$.
Note that, this neighbourhood is defined by similarity of inferential distributions, not by closeness of the tuning parameters $(h,r)$. As a consequence, it may reveal regions of the specification surface that are distributionally homogeneous even when the corresponding specifications are not adjacent on the grid. In this case, by choosing $\varepsilon=0.2568$ (i.e., the 25\% of the closest distributional specifications to the barycenter), the local set contains $|B_d(\overline{Q}_T,0.2568)|=96$ specifications only, with a local diameter of $R_{T|\varepsilon} = 0.2303$, a very low value if compared to the overall diameter. Finally, we do not report the specification curve for this example. Since the specification set is two-dimensional and approximately continuous, any one-dimensional ordering of specifications would be partly arbitrary. For instance, ordering specifications by their Hellinger distance from the Fréchet barycenter would make the distance panel mechanically monotone, adding little beyond the geometric information already shown in Figure~\ref{fig3}.

\section{Real case study}\label{sec4}

In this section, we describe the application of the distributional-geometric multiverse analysis to the well-known \textit{hurricane-name} case study \cite{hurricane}.\\

\noindent\textit{Data}. The dataset contains information on hurricanes, including the year of occurrence (\texttt{year}), the femininity score of the hurricane name (\texttt{masfem}), the minimum pressure (\texttt{min}), the highest wind speed (\texttt{wind}), the normalized amount of damage in 2015 dollars (\texttt{ndam15}), and the total number of deaths (\texttt{alldeaths}). The scientific target is the association between the femininity of the hurricane's name and the expected number of deaths, after adjustment for hurricane-related confounders. For each hurricane $i=1,\ldots,n$, let $Y_i$ denote the total number of deaths. Since the outcome is a count variable and is characterized by right skewness and overdispersion, we use a quasi-Poisson generalized linear model with log link. For a given specification $\sigma$, the linear predictor of the model is
$$\log(\mu_i) = \alpha + \beta \mathrm{masfem}_i + \gamma_1 f_1(\mathrm{year}_i)+ \gamma_2 f_2(\mathrm{min}_i)+ \gamma_3 f_3(\mathrm{wind}_i)+ \gamma_4 f_4(\mathrm{ndam15}_i),$$
with $\beta \in \mathbb R$ being the parameter of interest. This represents the log-multiplicative association between the femininity score of the hurricane's name \texttt{masfem} and the expected number of deaths, conditional on the selected transformations of the adjustment variables indicated as $f_1, \ldots, f_4$. \\

\noindent\textit{Specification set}. The multiverse is generated including all four confounding factors in the model and by varying the transformations applied to them. For each variable, we consider four possible transformations: identity, logarithm, square root, and a second-order polynomial. Thus, each specification corresponds to one combination of transformations, that is $\sigma = (f_1, f_2,f_3,f_4)$, and the admissible specification set contains $|\Sigma^\ast| = 4^4 = 256$ model specifications.\\

\noindent\textit{Point-valued and distributional multiverse}. For each specification $\sigma \in \Sigma^\ast$, let $\hat\beta_\sigma$ be the estimated coefficient of \texttt{masfem}, and let $\mathrm{\hat{se}}(\beta_\sigma)$ be its standard error. The point-valued multiverse is therefore

$$\mathcal T(\mathbf y)=\{ \hat\beta_\sigma : \sigma \in \Sigma^\ast\},$$
whereas the distributional multiverse $\mathcal Q_T(\mathbf y)$ is given by the collection of each asymptotic Gaussian sampling distribution $\mathcal N(\hat\beta_\sigma,\mathrm{\hat{se}}(\hat\beta_\sigma)^2)$. \\

\noindent\textit{Multiverse geometry}. To make $\mathcal Q_T(\mathbf y)$ metric, we use the squared Wasserstein distance, which is equivalent here to the Euclidean geometry between points in the $(\hat\beta_\sigma,\mathrm{\hat{se}}({\beta}_\sigma))$ plane. Another possible choice is provided by the Hellinger distance, which instead measures the discrepancy between distributions through their degree of overlap. For two probability distributions $Q_{\sigma,T}=\mathcal N(m_\sigma,s_\sigma^2)$ and $Q_{\sigma',T}=\mathcal N(m_{\sigma'},s_{\sigma'}^2)$, where $s_\sigma$ and $s_{\sigma'}$ denote standard errors, the Hellinger distance can be conveniently expressed as follows:
$$
H(Q_{\sigma,T},Q_{\sigma',T})
=
\left[
1-
\sqrt{\frac{2s_\sigma s_{\sigma'}}{s_\sigma^2+s_{\sigma'}^2}}
\exp\left\{
-\frac{(m_\sigma-m_{\sigma'})^2}{4(s_\sigma^2+s_{\sigma'}^2)}
\right\}
\right]^{1/2}.
$$
Thus, the same collection of specification-wise inferential distributions can be studied under different notions of proximity, each emphasizing a different aspect of the multiverse geometry.\\

\noindent\textit{Results}. The Wasserstein diameter is $\Delta_{T,W_2}(\mathbf y)=0.1183$. The two specifications attaining the diameter are reported in Table~\ref{tab2}. They differ mainly in the estimated location of the inferential distribution for the $\mathrm{masfem}$ effect, with estimated coefficients equal to $0.0877$ and $0.2047$, respectively. Hence, the diameter is driven mainly by differences in the location of the inferential distributions, although differences in uncertainty also contribute.
The unweighted Wasserstein barycenter is $\bar Q_{T,W_2}=\mathcal N(0.1492,0.0887^2)$, corresponding to a multiplicative effect $\exp(0.1492)=1.161$. The barycenter is not one of the fitted models, rather it represents the geometric center of the distributional multiverse under equal weights. It provides a compact summary of the central tendency of the multiverse while preserving an uncertainty component.

\begin{table}[!h]
	\centering
	\caption{Specifications attaining the diameter in the hurricane-name multiverse in both Wasserstein and Hellinger geometries.}
	\label{tab2}
	\begin{tabular}{l|rllllrrr}
		\toprule
		\textbf{Distance} & $\sigma$ & $f_{\mathrm{year}}$ & $f_{\mathrm{min}}$ & $f_{\mathrm{wind}}$ & $f_{\mathrm{ndam15}}$ & $\hat\beta_\sigma$ & $\mathrm{\hat{se}}(\beta_\sigma)$ & $\exp(\hat\beta_\sigma)$ \\
		\midrule
		\multirow{2}{*}{Wasserstein}
		& 126 & $\log$ & $\mathrm{poly2}$ & $\mathrm{poly2}$ & $\log$ & 0.0877 & 0.1022 & 1.0917 \\
		& 84  & $\mathrm{poly2}$ & identity & $\log$ & $\log$ & 0.2047 & 0.0845 & 1.2272 \\
		\midrule
		\multirow{2}{*}{Hellinger}
		& 128 & $\mathrm{poly2}$ & $\mathrm{poly2}$ & $\mathrm{poly2}$ & $\log$ & 0.0908 & 0.0851 & 1.0951 \\
		& 84 & $\mathrm{poly2}$ & identity & $\log$ & $\log$ & 0.2047 & 0.0845 & 1.2272 \\
		\bottomrule
	\end{tabular}
\end{table}

Under the Hellinger geometry, the distributional diameter is $\Delta_{T,H}(\mathbf y)=0.4493$. The two specifications attaining the Hellinger diameter are reported in Table~\ref{tab2}. Recall that the Hellinger diameter is bounded between 0 and 1 and measures the maximum lack of overlap between specification-wise inferential distributions. Therefore, the pair of specifications attaining the Hellinger diameter should be interpreted as the two admissible models whose Gaussian approximations are most separated in terms of distributional overlap.
The unweighted Hellinger barycenter is $\bar Q_{T,H}=\mathcal N(0.1497,0.0912^2)$, corresponding to a multiplicative effect $\exp(0.1497)=1.1615$. 

Figure~\ref{fig3} displays the hurricane multiverse under the two geometries. Panel (A) shows the effect--uncertainty representation together with the Wasserstein barycenter. The horizontal axis captures variation in the estimated $\texttt{masfem}$ effect, whereas the vertical axis captures variation in the corresponding standard error. Panel (B) shows the same specification-wise inferential distributions together with the Hellinger barycenter through a multidimensional scaling projection of the Hellinger distance matrix. Therefore, the axes do not correspond to the estimated effect and its standard error, but to low-dimensional coordinates preserving, as much as possible, the pairwise Hellinger distances among specification-wise inferential distributions.
\begin{figure}[t]
	\centering
	\includegraphics[width=\textwidth]{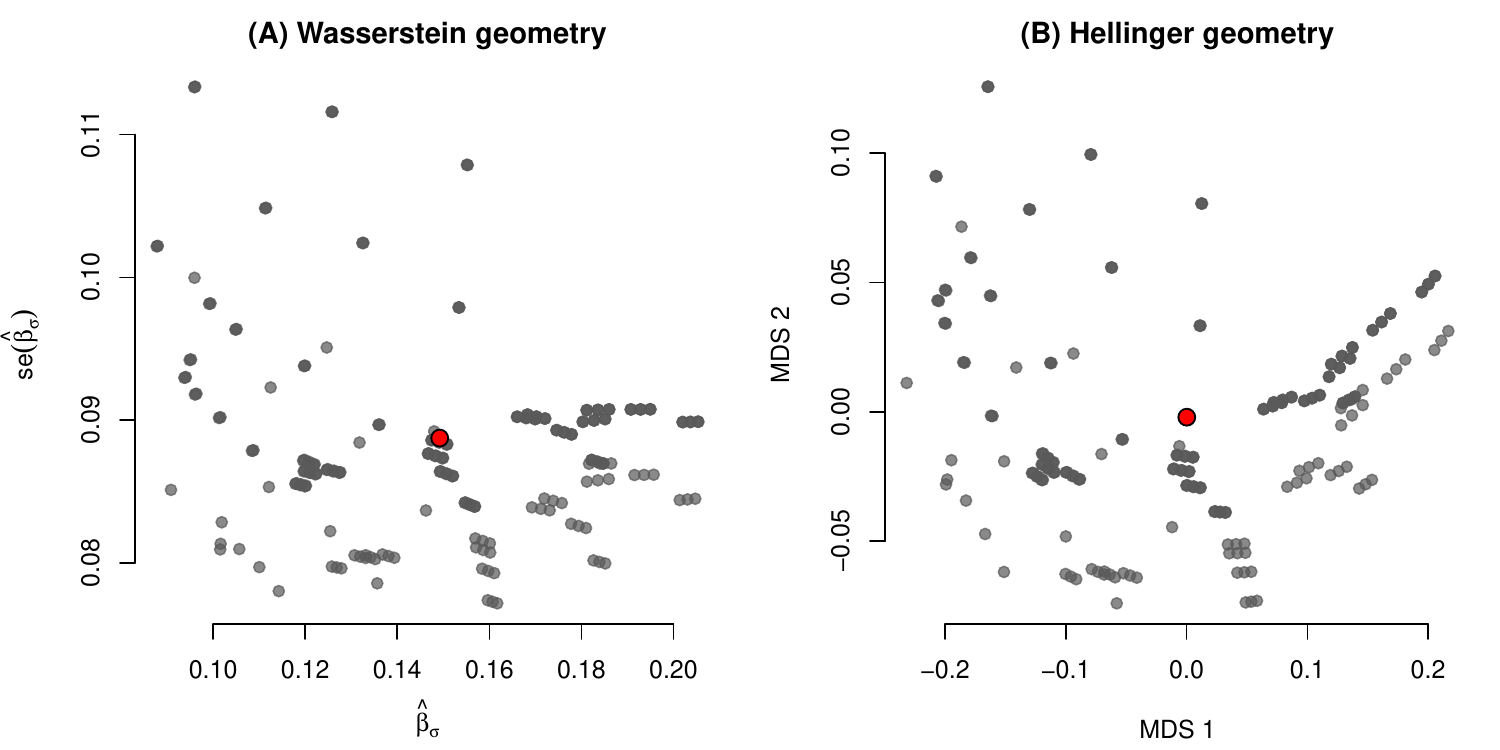}
	\caption{
		Distributional-geometric representation of the hurricane-name multiverse under two alternative geometries.
		Panel (A) displays the Wasserstein geometry in the effect--uncertainty plane, where each point represents one admissible specification located according to the estimated coefficient $\hat\beta_\sigma$ of $\mathrm{masfem}$ and its standard error $\mathrm{\hat{se}}(\beta_\sigma)$; the red point denotes the Wasserstein barycenter.
		Panel (B) displays a multidimensional scaling (MDS) representation of the Hellinger distance matrix, where point locations reflect pairwise Hellinger dissimilarities among the specification-wise inferential distributions; the red point denotes the corresponding Hellinger barycenter projected in the same MDS space.
	}
	\label{fig3}
\end{figure}
We can notice how the admissible specifications do not form a single homogeneous cloud. Rather, they occupy distinct regions of the plane, indicating that some modelling choices affect mainly the estimated effect, while others also affect its uncertainty.
Indeed, the geometric representation also helps identify which analytical decisions drive the multiverse geometry. In Fig.~\ref{fig4}, we represent in colour each specification as a point $(\hat\beta_\sigma,\mathrm{\hat{se}}(\beta_\sigma))$ according to the transformation used for each adjustment variable. This allows us to inspect whether a given modelling choice mainly affects the estimated coefficient, its uncertainty, or both. In this example, the transformations of $\texttt{wind}$ and $\texttt{ndam15}$ are particularly relevant for the right-hand region of the multiverse, where the estimated association between $\texttt{masfem}$ and deaths is larger. This diagnostic is meant to complement a specification curve by showing how modelling choices are arranged in the effect--uncertainty plane.\\
\begin{figure}[!h]
	\centering
	\includegraphics[width=\textwidth]{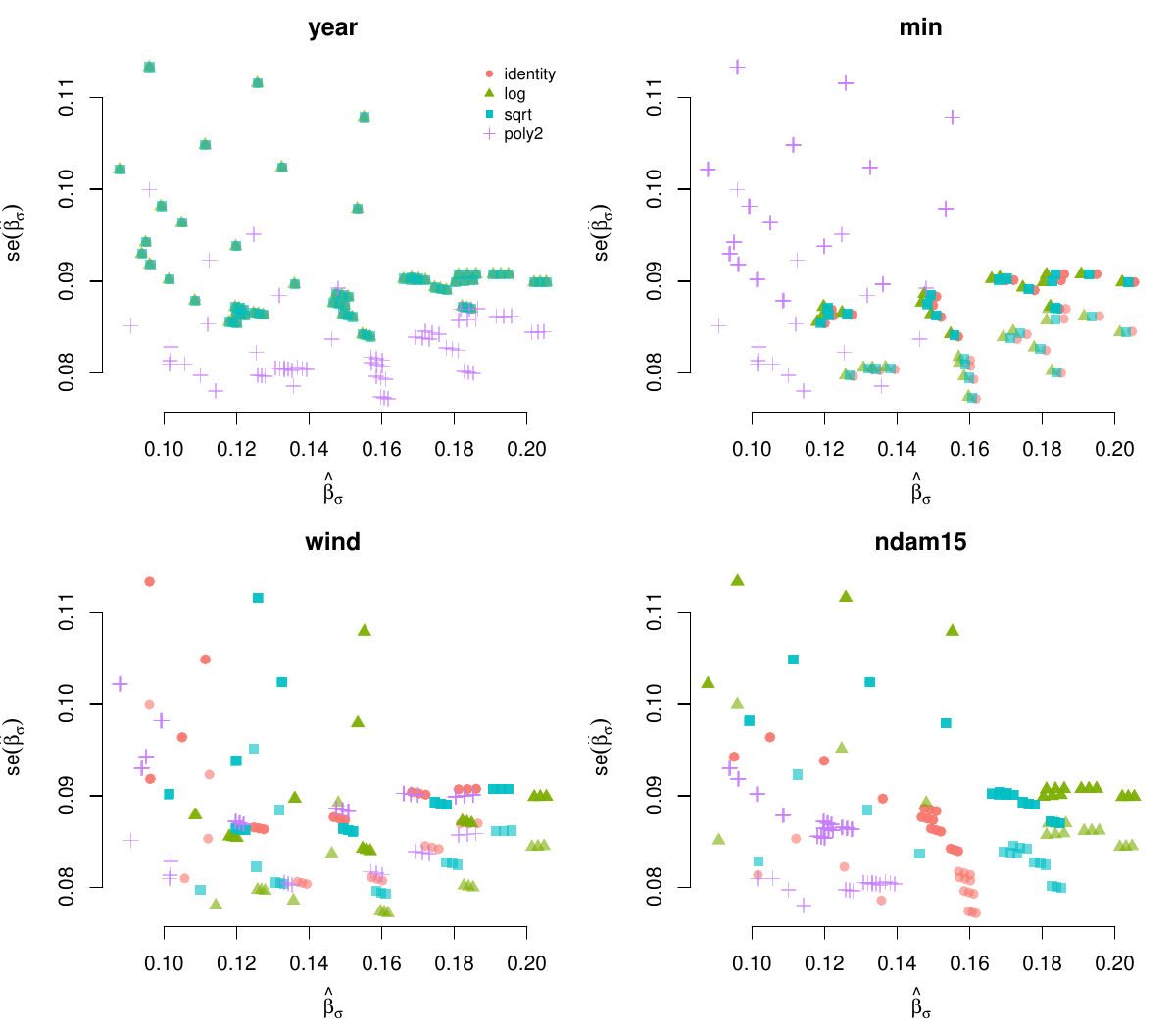}
	\caption{Distributional-geometric representation of the hurricane-name multiverse. Each point/specification is coloured according to the transformation used for each adjustment variable: \texttt{year}, \texttt{min}, \texttt{wind}, and \texttt{ndam15}. Each panel considers one variable at a time, with the colour and shape indicating which transformation was applied to the variable displayed in the panel.}
	\label{fig4}
\end{figure}

\begin{figure}[!h]
	\centering
	\includegraphics[width=\textwidth]{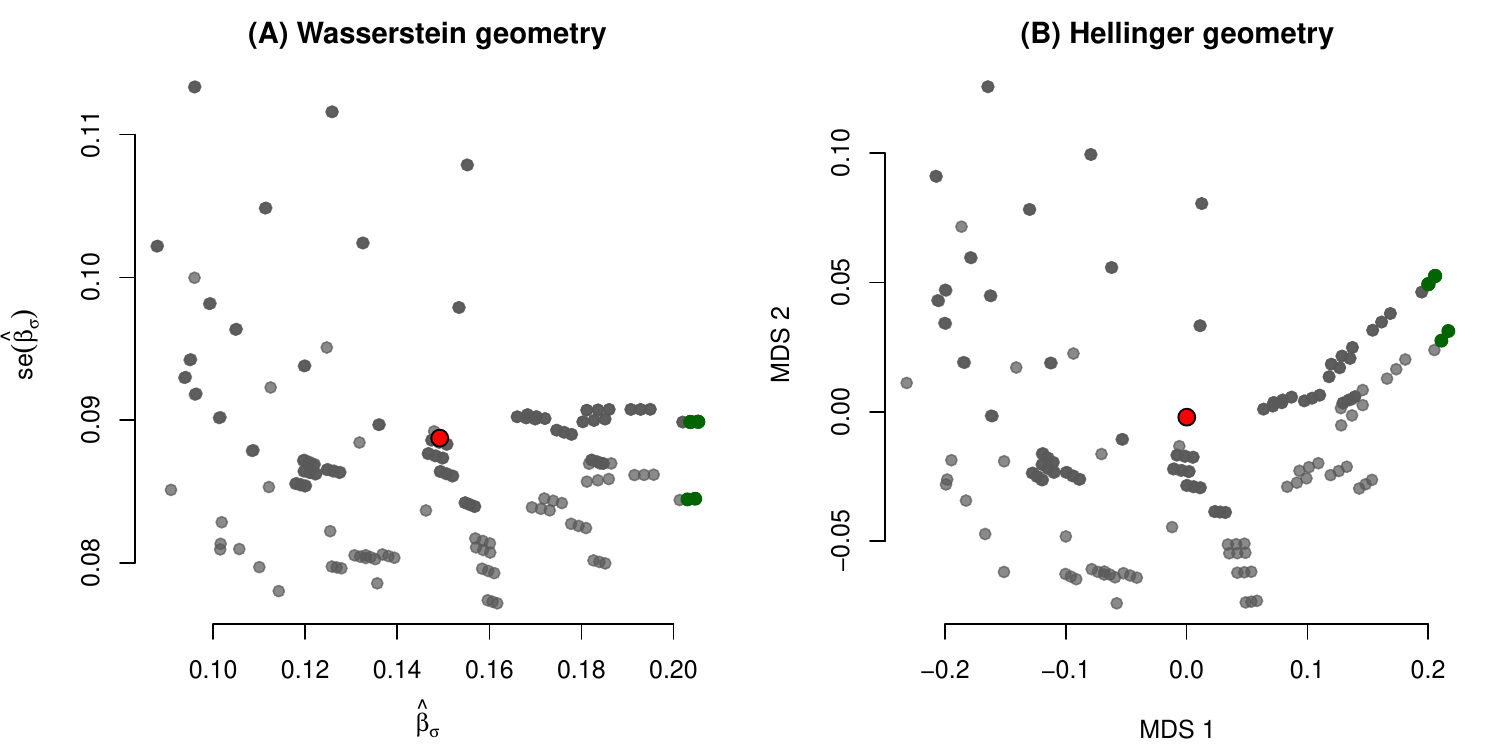}
	\caption{
		Distributional-geometric representation of the hurricane-name multiverse with PIMA-selected specifications highlighted. 
		Panel (A) displays the Wasserstein geometry in the effect--uncertainty plane, where each point represents one admissible specification located according to \(\hat\beta_\sigma\) and \(\mathrm{\hat{se}}(\beta_\sigma)\). 
		Panel (B) displays a multidimensional scaling representation of the Hellinger distance matrix, where point locations reflect pairwise Hellinger dissimilarities among the specification-wise inferential distributions. 
		In both panels, green points identify specifications that remain significant after PIMA adjustment, and the red point denotes the corresponding barycenters.
	}
	\label{fig5}
\end{figure}

\noindent\textit{Further analyses}. With the aim of integrating an inferential-oriented multiverse results into the general distributional-geometric multiverse analysis, we re-analysed the \textit{hurrican-names} data using PIMA \cite{girardi2024}.\footnote{Although other alternatives might be considered for this purpose, we opted for PIMA because it provides a post-selection permutation-based test, which controls for the type I error rate by ensuring FWER control globally.} More in detail, PIMA tests the null hypothesis $H_{0,\sigma}:\beta_\sigma=0$ for each specification $\sigma\in\Sigma^\ast$, and it also provides a global test across the multiverse together with adjusted $p$-values for the model-specific tests. In the current study, the global PIMA test gives $p=0.0472$ (with $5000$ sign flips) and, after PIMA adjustment at level $\alpha=0.05$, only eight specifications remain significant (see Table~\ref{tab3}). 
\begin{table}[t]
	\centering
	\small
	\caption{Specifications remaining significant after PIMA adjustment in the hurricane-name multiverse.}
	\label{tab3}
	\begin{tabular}{rllllrrrr}
		\toprule
		\(\sigma\) 
		& \(f_{\mathrm{year}}\) 
		& \(f_{\mathrm{min}}\) 
		& \(f_{\mathrm{wind}}\) 
		& \(f_{\mathrm{ndam15}}\) 
		& \(\hat\beta_\sigma\) 
		& \(\mathrm{\hat{se}}(\beta_\sigma)\) 
		& \(\exp(\hat\beta_\sigma)\) 
		& {\footnotesize \(W_2(Q_{\sigma,T},\bar Q_{T,W_2})\)} \\
		\midrule
		92 & poly2    & sqrt     & log & log & 0.2031 & 0.0844 & 1.2251 & 0.0540 \\
		90 & log      & sqrt     & log & log & 0.2036 & 0.0899 & 1.2259 & 0.0544 \\
		91 & sqrt     & sqrt     & log & log & 0.2037 & 0.0899 & 1.2259 & 0.0545 \\
		89 & identity & sqrt     & log & log & 0.2037 & 0.0899 & 1.2260 & 0.0545 \\
		84 & poly2    & identity & log & log & 0.2047 & 0.0845 & 1.2272 & 0.0557 \\
		82 & log      & identity & log & log & 0.2053 & 0.0899 & 1.2279 & 0.0561 \\
		83 & sqrt     & identity & log & log & 0.2054 & 0.0899 & 1.2280 & 0.0562 \\
		81 & identity & identity & log & log & 0.2054 & 0.0899 & 1.2280 & 0.0562 \\
		\bottomrule
	\end{tabular}
\end{table}
We can notice how they form a well-structured subset of the multiverse of analyses. Indeed, all the selected specifications use \(\log(\texttt{wind})\) and \(\log(\texttt{ndam15})\), while the transformation of \(\texttt{min}\) alternates between identity and square root. By contrast, all four transformations of \(\texttt{year}\) appear among the selected specifications. This suggests that the PIMA-adjusted significant results are not uniformly spread across the multiverse, but are concentrated in a specific region of the analytical decision space, with estimated coefficients around $0.203$--$0.205$. To better understand this result, Figure~\ref{fig5} represents the PIMA results into the distributional multiverse, displaying in green the points which identify the specifications that remain significant after adjustment. Interestingly, the geometric representation makes explicit both where the PIMA-selected specifications lie in the distributional multiverse (particularly, in the frontier of the multiverse) and which modelling choices characterize that region (see Table \ref{tab3}).

\section{Conclusions}\label{sec5}

This paper has proposed a distributional-geometric view of multiverse analysis. The starting point is simple: a multiverse does not only generate a collection of estimates, tests, or model labels. On the contrary, each admissible specification also carries an inferential distribution for the target quantity. Treating these specification-wise distributions as the primary objects makes it possible to study the multiverse as a configuration in a space of probability measures.

To the best of our knowledge, this is the first explicit distributional-geometric formulation of multiverse analysis. The proposal is not meant to replace current multiverse analysis tools such as specification curves, model weights, or post-selection techniques. Rather, it adds a further descriptive layer, providing a unified and consistent modeling approach that can enhance the understanding and the interpretation of multiverse analysis results. Moreover, once a distance between inferential distributions is chosen, one can study the multiverse analysis from a geometric perspective, by asking which specifications are close, which ones are peripheral, how heterogeneous the multiverse is, and where a barycentric summary lies relative to the full set of admissible analyses. In this sense, the framework gives a formal language for describing features of a multiverse that are not visible from scalar summaries alone.

The present contribution should therefore be read as a first descriptive and representational step, rather than as a calibrated post-selection procedure. Future work may develop this inferential layer in at least two directions. A first direction concerns decision rules based on the geometry of $\mathcal Q_T(\mathbf y)$, for example procedures that select, aggregate, or report subsets of specifications while controlling type I error or family-wise error rates, in the spirit of post-selection frameworks such as PIMA \cite{girardi2024}. A second direction concerns the asymptotic behaviour of the distributional representation itself. Under suitable regularity conditions, one may ask whether the estimated specification-wise distributions $Q_{\sigma,T}$ converge, possibly uniformly over admissible or correctly specified subsets of $\Sigma^\ast$, to their target inferential distributions, and whether geometric summaries such as diameters, barycentres, and dispersions converge to well-defined population counterparts. We leave these calibration and convergence questions to future work.

\clearpage
\bibliographystyle{plain}
\bibliography{biblio}

\end{document}